\documentclass[traditabstract,rnote]{aa}
\usepackage[utf8]{inputenc}
\usepackage{amsmath}
\usepackage{amssymb}
\usepackage{microtype}
\usepackage{natbib}
\usepackage{graphicx}
\usepackage{txfonts}
\usepackage{color}
\usepackage{multirow}
\usepackage{xspace}
\bibpunct{(}{)}{;}{a}{}{,} 
\usepackage[plainpages=false]{hyperref}

\newcommand{\src}{\object{V~0332$+$53}\xspace}

\title{Orbital parameters of \src from 2015 giant outburst data.}
\author{V.\,Doroshenko\inst{1}, S.\,Tsygankov\inst{2}, A.\,Santangelo\inst{1}}	
\institute{Institut für Astronomie und Astrophysik, Sand 1, 72076 Tübingen, Germany\and
Tuorla Observatory, Department of Physics and Astronomy, University of Turku, V\"ais\"al\"antie 20, FI-21500 Piikki\"o, Finland}

\begin{document}

\bibliographystyle{aa}

\abstract{We present the updated orbital solution for the transient Be X-ray
binary \src complementing historical measurements with the data from the
gamma-ray burst monitor onboard \emph{Fermi} obtained during the outburst in
June-October 2015. We model the observed changes in the spin-frequency of the
pulsar and deduce the orbital parameters of the system. We significantly
improve existing constrains and show that contrary to the previous findings no
change in orbital parameters is required to explain the spin evolution of the
source during the outbursts in 1983, 2005 and 2015. The reconstructed intrinsic
spin-up of the neutron star during the latest outburst is found to be
comparable with previosly observed values and predictions of the accretion
torque theory.}

\authorrunning{V. Doroshenko et al.}
\maketitle

\section{Introduction} The bright X-ray transient source \src was discovered by
the Vela~5B satellite during an outburst in 1973 \citep{Terrel84}. Observations
with EXOSAT ten years later allowed the position of the source to be determined
the X-ray pulsations with period of $\sim4.4$\,s \citep{Stella85} modulated by
motion in an eccentric orbit ($e\sim0.3$) with period of $\sim34.3$\,d to be
detected. The optical counterpart of the X-ray pulsar has been identified to be
a Be star \citep{Honeycutt85} at a distance of $\sim7$\,kpc \citep{Negu99}.

Typically for Be systems, \src exhibits two types of X-ray outbursts with peak
luminosities of $\sim10^{38}\,{\rm erg\,s}^{-1}$ and $\sim10^{37}\,{\rm
erg\,s}^{-1}$ associated with enhanced accretion onto the neutron star from the
Be circumstellar disk close to the periastron. The so-called ``normal'' or
Type~I outburst occur regularly as the neutron star passes through the Be disk
and have comparatively low luminosity. Longer and more luminous type II or
``giant'' outbursts are more rare and thought to be related to the formation of
a stable accretion disk around the neutron star during a periastron passage
\citep{Okazaki02,Martin14}. Four type II outbursts and a number of type I
outbursts have been detected from the source so far
\citep{Terrel84,Tsunemi89,Swank04,Camero15,Myatel,Nakajima15}.

In this note we report on the analysis of timing properties of \src during the
most recent type II outburst which took place in June-October~2015 together
with older observations and provide the updated orbital solution for the system.

\section{Data analysis and results} Orbital parameters of \src were first
determined by \cite{Stella85} using the EXOSAT data. Later \cite{Zhang05} and
\cite{Raichur10} used Rossi X-Ray Timing Explorer (RXTE) and International
Gamma-Ray Astrophysics Laboratory (INTEGRAL) observations performed during the
2004-2005 giant outburst to refine the orbital solution. These authors found
significantly larger projected semimajor axis $a\sin{i}$ value than reported by
\cite{Stella85} which was associated with the apsidal motion in the system.

The most recent outburst in 2015 has been monitored with several instruments.
However, the best timing information is provided by the Gamma-ray burst monitor
(GBM) on board \emph{Fermi} \citep{Meegan} which regularly measured spin
frequency of the source every 1-3 days throghout the outburst. The GBM consists
of 12 \emph{NaI} and two \emph{BGO} detectors providing the full-sky coverage
in 8~\,keV$-$40\,MeV energy range and is designed to detect and localize gamma
ray bursts. However, it also proved to be very useful for detection and
monitoring of pulsed signals from X-ray pulsars. In particular,
the \emph{Fermi}~GBM team regularly
publishes\footnote{http://gammaray.msfc.nasa.gov/gbm/science/pulsars.html} spin
histories of selected pulsars including \src based on the analysis of
\emph{CTIME} data from \emph{NaI} detectors in 8-50\,keV energy range.

\begin{figure*}[!ht]
	\centering
		\includegraphics[width=0.33\textwidth]{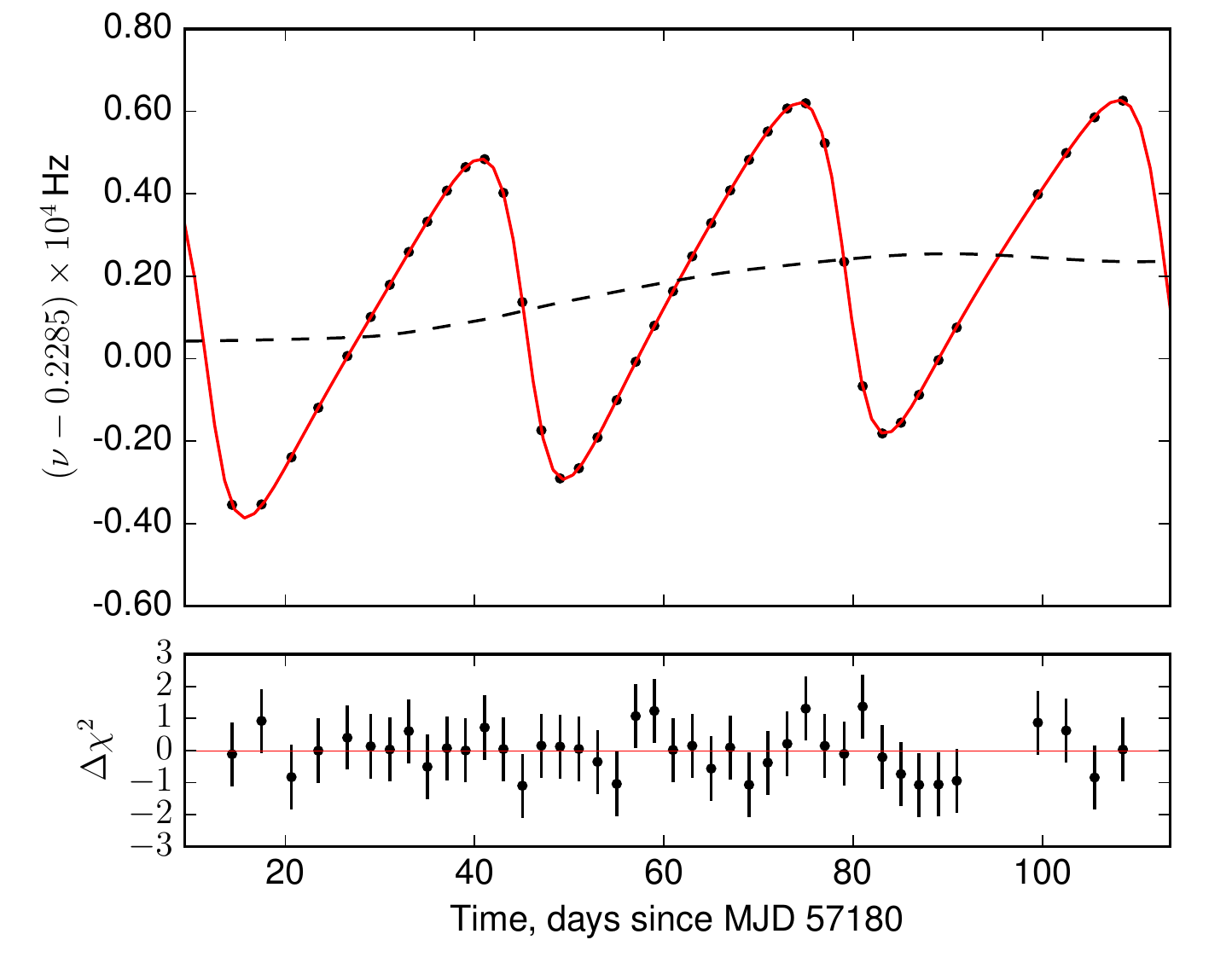}
		\includegraphics[width=0.33\textwidth]{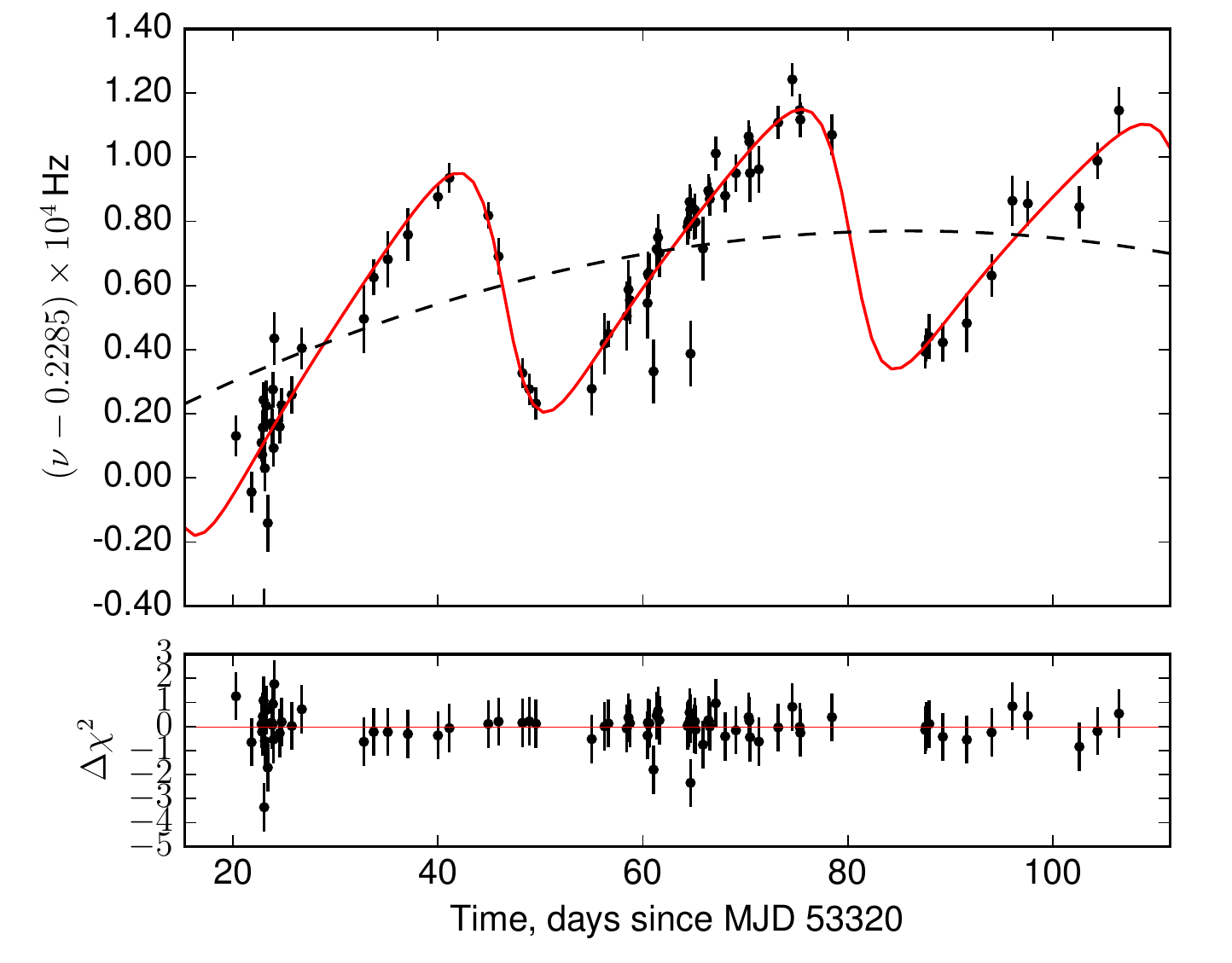}
		\includegraphics[width=0.33\textwidth]{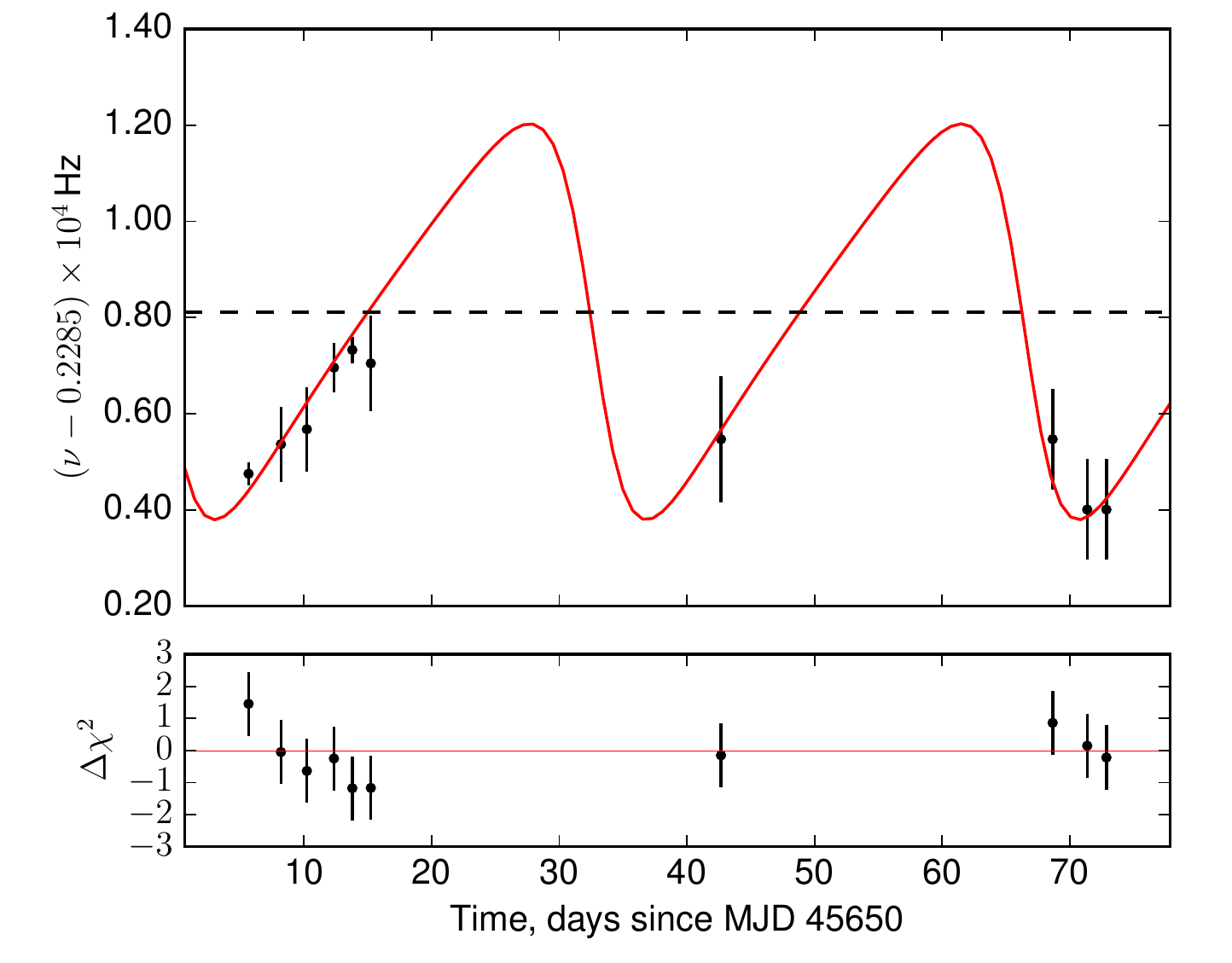}
	\caption{Observed pulse frequency modulated by orbital motion as measured with (left to right) \emph{Fermi}~GBM, RXTE and EXOSAT and Tenma during the three major outbursts from the source. Reconstructed intrinsic pulsar frequency (black dashed line) modulated by motion along the orbit with best-fit parameters (red line) together with fit residuals are also shown.}
	\label{fig:gbm}
\end{figure*}

Pulsations from \src were first detected by GBM on MJD~57194 and the data used
in this publication covers the interval until MJD~57290, i.e. almost three full
orbital cycles. The orbital modulation of the spin frequency superimposed by a
spin-up trend is clearly visible in the raw data shown in Fig.~\ref{fig:gbm}.
We use these measurements to constrain the orbital parameters and intrinsic
spin evolution of the pulsar using the same approach as \cite{Zhang05}. In
addition, we repeat the analysis of RXTE data carried out by \cite{Raichur10}.
In particular, we reconstruct the spin history of the pulsar during the giant
outburst in 2005 using the epoch-folding period search and RXTE PCA lightcurves
in 3-21\,keV energy range between MJD~53332 and 53432. Using the obtained spin
frequency measurements we followed procedures described in \cite{Zhang05} and
\cite{Raichur10} to constrain the orbital parameters of the system. We modeled
simultaneously the data from 2005 and 2015 outbursts together with historical
pulse frequency measurements reported by \cite{Stella85} and \cite{Makishima}
for the outburst in 1983-1984.

We emphasize that both the orbital modulation and intrinsic variation of the
pulsars spin frequency must be modeled in order to constrain the orbital
parameters of the system. The intrinsic spin evolution of \src during an
outburst is rather complicated as found by \cite{Zhang05,Raichur10} who had to
include the second pulse period derivative to describe it adequately. This is
to be expected as the accretion torque spinning up the neutron star depends on
the accretion rate which changes dramatically and also true for the latest
outburst. In fact, we were unable to describe the spin history measured with
\emph{Fermi}~GBM even with inclusion of the third spin frequency derivative as
a parameter.

Therefore, we opted to model the intrinsic spin evolution of the pulsar as a
smooth interpolating function (i.e. using the piecewise cubic hermite
interpolating polynomial) defined by frequency values at fixed times $T_i =
{\rm MJD} 57194,57208,57220,57230,57250,57270,57290$. Following \cite{Zhang05}
we included a systematic uncertainty of $7\times10^{-8}$\,Hz and
$7\times10^{-6}$\,Hz for \emph{Fermi}~GBM and RXTE data respectively.

Joint fitting of all three datasets results in $\chi^2_{red}=1.02$ for 115
degrees of freedom and the best-fit results are presented in Fig.~\ref{fig:gbm}
and Table.~\ref{tab:orpar}. We wish to emphasize that contrary to findings by
\cite{Zhang05} it was possible to describe the spin evolution of the source
during all three outbursts without assuming any change in orbital parameters.
Obtained results are generally consistent with the values reported by
\cite{Zhang05} and \cite{Raichur10} with the exception of the orbital period
value. We note that in our case it is much better defined as the periastron
passage time is very well constrained in both RXTE and \emph{Fermi} datasets
and the number of orbital cycles between the two outbursts is known. Other
discrepancies likely arise due to the different assumptions on the intrinsic
spin frequency of the pulsar coupled with insufficient orbital phase coverage
and limited accuracy of individual spin frequency measurements in previous
studies.

\paragraph{Implications for the binary orbit.} We have shown that it is
possible to describe the data outbursts in 1983, 2005 and 2015 assuming no
change in the orbital parameters, thus resolving the discrepancy in $a\sin(i)$
values reported by \cite{Stella85} and \cite{Zhang05}. The discussion by
\cite{Zhang05} regarding a possible apsidal motion in the system is thus not
required anymore. We conclude, therefore, that there is no evidence for apsidal
motion in the system.

On the other hand, the low $a\sin{i}$ value reported by \cite{Stella85} led
\cite{Negu99} to conclude that the Be companion in \src must be strongly
undermassive. Indeed, assuming $a\sin{i}\sim50$\,lt\,s the projected rotational
velocity of the Be star measured by \cite{Negu99}
$\upsilon_*\sin{i_*}\sim100-200$\,km\,s$^{-1}$ implies
$i_*\ge12^\circ-24^\circ$ if the intrinsic rotational velocity of the star is
below the break-up value $\upsilon_*\le0.8\upsilon_{\rm
break-up}\sim480$\,km\,s$^{-1}$. For orbital parameters reported by
\cite{Stella85} this indeed implies a strongly undermassive optical companion
with $M_*\le5-7\,M_\odot$ unless its equatorial plane is misaligned with the
orbital plane of the system \citep{Negu99}. For larger $a\sin{i}\sim78$\,lt\,s
found in this work the same considerations considerations imply
$M_*\le8-50\,M_\odot$, i.e. compatible with the expected mass of a Be star. We
conclude, therefore, that there is also no evidence for a tilt between its
equatorial plane and orbital plane of the system.

\paragraph{Intrinsic spin evolution} Once the orbital parameters of the system
are determined it is possible to assess the intrinsic spin evolution of the
pulsar and thus probe the accretion torques acting onto the neutron star. To
estimate the time spin frequency derivative we correct the observed frequency
values for motion in the binary system using the ephemeris obtained above and
calculate the frequency derivative comparing the measurements in adjacent time
intervals (propagating the uncertanties). The average spin-up rate of
$\dot{P}_{spin}\sim5\times10^{-6}$\,s\,d$^{-1}$ is comparable with
$\dot{P}_{spin}\sim8\times10^{-6}$\,s\,d$^{-1}$ reported by \cite{Zhang05} and,
as discussed by these authors, is in rough agreement with the accretion torque
theory.

\begin{figure}[t!] \centering \includegraphics[width=0.5\textwidth]{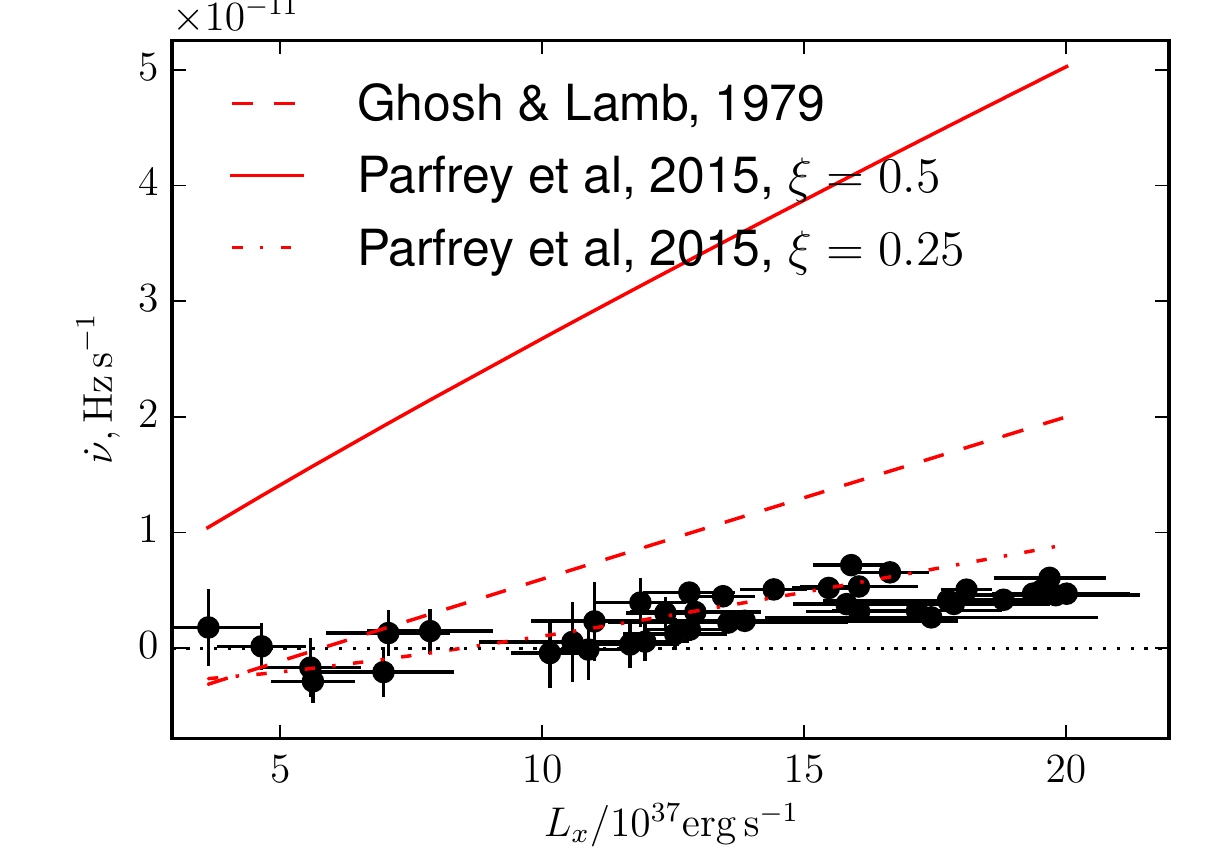}
\caption{Observed spin-up rate as function of source X-ray luminosity for the
rising (circles) and declining (diamonds) parts of the 2015 outburst. Model
predictions after \cite{Ghosh79} and \cite{Parfrey15} are also shown for reference.}
\label{fig:spinup} \end{figure}

The spin-up rate of the neutron star is expected to be correlated with the
accretion rate. Therefore, we estimate also the accretion rate for each time
interval where the spin-up rate was determined. In particular we multiply the
average \emph{Swift~BAT} count rate in 15-50\,keV range during respective time
interval by factor of $1.6\times10^{-7} {\rm erg\,s}^{-1}{\rm cm^{-2}}$ to
calculate the flux. This factor was calculated based on the comparison of the
\emph{Swift BAT} count rates and flux in 3-100\,keV range measured during the
pointed \emph{NuSTAR} observations on MJD~57223, 57275 and 57281 (detailed
analysis of pointed observations will be presented elsewhere). The luminosity
and accretion rate can then be estimated assuming the distance to the source of
7\,kpc and standard neutron star parameters. We find that the spin-up rate is
indeed correlated with the observed X-ray flux as illustrated in
Fig.~\ref{fig:spinup}.

Several accretion torque models describing the spin evolution of a neutron star
accreting from a disk have been proposed. The model by \cite{Ghosh79} invokes
angular momentum transport from a neutron star to an accretion disk threaded by
the magnetic field lines. As can be seen in Fig.~\ref{fig:spinup}, this model
predicts somewhat higher spin-up rate than observed (assuming the magnetic
field of the neutron star of $B=3.5\times10^{12}$\,G corresponding to the
highest cyclotron line energies reported by \citealt{Tsygankov06}). The
discrepancy is not large, however, and could be reconciled if we assume that
the accretion rate is overestimated by factor of two.

In a more recent study \cite{Parfrey15} argue that the angular momentum is
removed from the neutron star by the electromagnetic outflow along the field
lines opened by differential rotation of the magnetic field and the accretion
disk. The total torque in this model depends strongly on the magnetospheric
radius, which is assumed to constitute a fraction of the Alfvenic radius
$r_a=(\mu^4/2GM\dot{M}^2)^{1/7}$ for spherical accretion, i.e. $r_m=\xi r_a$.
The value of $\xi\sim0.5$ is expected from MHD simulations
\citep{Long05,Bessolaz08,Zanni13}. However, in this case the model predicts
significantly higher spin-up rate for \src than observed. To match the
observations with the model prediction one can assume either that the accretion
rate is over-estimated by factor of ten (which is unlikely) or that the $\xi$
must be reduced by factor of two (i.e. $\xi=0.25$ as presented in
Fig.~\ref{fig:spinup}) which could signify that the disk pushes further into
the magnetosphere than expected.

\begin{table}
	\begin{center}
	\begin{tabular}{llll}
		Parameter & This work & RP10 & Z05\\
		\hline
		$P_{orb}$, d & 33.850(1) &36.5(3) & 34.7(4) \\
		$a\sin(i)$, lt\,s &77.81(7) & 82.5(9) & 86(10) \\
		$e$ & 0.3713(8) & 0.417(7) &0.37(12)\\
		$\omega$, deg & 277.43(5) & 283.5(9) & 283(14)\\
		$T_{PA}$, MJD & 57157.88(3) & 53330.58(6) & 53367(1)\\
		$K_X$, km\,s$^{-1}$ & 53.97(5) & -- & 59(7) \\
		$f(M)$, $M_\odot$ & 0.441(1)  & -- & 0.58(23) \\
		\hline
	\end{tabular}
	\end{center}
	\caption{The best-fit orbital parameters of \src including the \emph{Fermi}~GBM data from 2015 outburst (left column). Values obtained by \cite{Raichur10} (RP10) and \cite{Zhang05} (Z05) are also shown for reference. All uncertainties are at $1\sigma$ confidence level.}
	\label{tab:orpar}
\end{table}

\section{Conclusions} In this note we analyzed the \emph{Fermi}~GBM spin
history of \src during the giant outburst in 2015 together with historical data
from previous outbursts in 1983-1984 and 2004-2005. Our results are generally
consistent with earlier estimates by \cite{Zhang05,Raichur10} and similar to
the preliminary findings by the \emph{Fermi}~GBM team (based only on the data
from the latest outburst).

For the first time we succeeded to describe the spin evolution of the source
during all three outbursts with no change in orbital parameters between them
thus resolving a long standing discrepancy between the orbital solutions
reported by \cite{Stella85}, \cite{Zhang05}, and \cite{Raichur10}. Our finding
makes the discussion by \cite{Zhang05} regarding the possible apsidal motion in
the system unnecessary. We note also that in the light of updated ephemeris the
conclusion by \cite{Negu99} regarding the misalignment of orbital plane of the
system and Be stars equatorial plane does not hold anymore. We were also able
to significanlty improve the accuracy of the orbital solution.

We find the the intrinsic spin evolution of the pulsar to be complicated with
the spin-up rate being correlated with the accretion rate. The observed spin-up
is qualitatively consistent with existing torque models assuming the neutron
star has the magnetic field of $\sim3\times10^{12}$\,G as determined from
cyclotron line energy although model uncertanties and the uncertainty in
accretion rate of the source prevent us from any conclusions regarding
preferred torque model.
 
\begin{acknowledgements}
This work is based on spin-histories provided by \emph{Fermi}~GBM pulsar
project. VD and AS thank the Deutsches Zentrums for Luft- und Raumfahrt (DLR)
and Deutsche Forschungsgemeinschaft (DFG) for financial support (grant
DLR~50~OR~0702). ST acknowledges support by the Russian Scientific Foundation
grant 14-12-01287.
\end{acknowledgements}
\bibliography{biblio}	\vspace{-0.3cm}\end{document}